\documentclass[twocolumn,astrosymb]{aastex631}

\usepackage{xspace}
\usepackage{tabularx}
\usepackage{booktabs}
\usepackage{threeparttable}
\usepackage{longtable}

\newcommand{\rev}[1]{{#1}\xspace}

\newcommand{\pyttv}{\texttt{PyTTV}\xspace}
\newcommand{\pytransit}{\texttt{PyTransit}\xspace}
\newcommand{\rebound}{\texttt{Rebound}\xspace}
\newcommand{\reboundx}{\texttt{Reboundx}\xspace}
\newcommand{\emcee}{\texttt{emcee}\xspace}
\newcommand{\tess}{TESS\xspace}

\newcommand{\mjup}{\ensuremath{M_\mathrm{Jup}}\xspace}
\newcommand{\rjup}{\ensuremath{R_\mathrm{Jup}}\xspace}
\newcommand{\msun}{\ensuremath{M_\odot}\xspace}
\newcommand{\rsun}{\ensuremath{R_\odot}\xspace}
\newcommand{\rearth}{\ensuremath{R_\oplus}\xspace}
\newcommand{\mearth}{\ensuremath{M_\oplus}\xspace}
\newcommand{\smass}{\ensuremath{M_\star}\xspace}
\newcommand{\sradius}{\ensuremath{R_\star}\xspace}

\newcommand{\p}{\ensuremath{P}\xspace}
\newcommand{\epoch}{\ensuremath{T_0}\xspace}

\newcommand{\rpoverrstar}{\ensuremath{R_\mathrm{p}/R_\star}\xspace}
\newcommand{\gimp}{\ensuremath{g}\xspace}

\newcommand{\vsini}{\ensuremath{V\sin{i_\star}}\xspace}
\newcommand{\teff}{\ensuremath{T_{\mathrm{eff}}}\xspace}
\newcommand{\feh}{[Fe/H]\xspace}
\newcommand{\logg}{\ensuremath{\log g}\xspace}
\newcommand{\mua}{\ensuremath{\rm\mu_{\alpha\star}}\xspace}
\newcommand{\mud}{\ensuremath{\rm\mu_{\delta\star}}\xspace}
\newcommand{\lum}{\ensuremath{L_\star}\xspace}

\newcommand{\kms}{km\,$\mathrm{s}^{-1}$}
\newcommand{\msunit}{m\,$\mathrm{s}^{-1}$}
\newcommand{\my}{mas\,$\mathrm{yr}^{-1}$}

\newcommand{\UP}[1]{\ensuremath{\mathcal{U}(#1)}\xspace}
\newcommand{\NP}[1]{\ensuremath{\mathcal{N}(#1)}\xspace}

\newcommand{\planetb}{TOI-1408~b\xspace}
\newcommand{\planetc}{TOI-1408~c\xspace}
\newcommand{\host}{TOI-1408\xspace}
\newcommand{\periodb}{\ensuremath{4.42}\,d\xspace}

\newcommand{\radiusb}{\ensuremath{2.4\pm0.5}\rjup}
\newcommand{\radiusc}{\ensuremath{2.22\pm0.06}\rearth}
\newcommand{\massb}{\ensuremath{1.86\pm0.02}\mjup}
\newcommand{\massc}{\ensuremath{7.6\pm0.2}\mearth}
\newcommand{\rot}{\ensuremath{7.9\pm0.6}\,d\xspace}

\begin{document}

\title{TOI-1408: Discovery and Photodynamical Modeling of a Small Inner Companion to a Hot Jupiter Revealed by TTVs}

\correspondingauthor{Judith Korth}
\email{judithkorth@googlemail.com}

\author[0000-0002-0076-6239]{Judith Korth}
\affiliation{Lund Observatory, Division of Astrophysics, Department of Physics, Lund University, Box 118, 22100 Lund, Sweden}

\author[0000-0002-1887-1192]{Priyanka Chaturvedi}
\affiliation{Th\"uringer Landessternwarte Tautenburg, Sternwarte 5, 07778 Tautenburg, Germany}
\affiliation{Tata Institute of Fundamental Research (TIFR), Homi Bhabha Road, Colaba, Mumbai - 400005, India}

\author[0000-0001-5519-1391]{Hannu Parviainen}
\affiliation{Departamento de Astrofísica, Universidad de La Laguna (ULL), E-38206 La Laguna, Tenerife, Spain}
\affiliation{Instituto de Astrofísica de Canarias (IAC), E-38205 La Laguna, Tenerife, Spain}

\author{Ilaria Carleo}
\affiliation{INAF -- Osservatorio Astrofisico di Torino, Via Osservatorio 20, I-10025, Pino Torinese, Italy}
\affiliation{Instituto de Astrofísica de Canarias (IAC), E-38205 La Laguna, Tenerife, Spain}
\affiliation{Departamento de Astrofísica, Universidad de La Laguna (ULL), E-38206 La Laguna, Tenerife, Spain}

\author{Michael Endl}
\affiliation{McDonald Observatory, The University of Texas, Austin Texas USA}
\affiliation{Center for Planetary Systems Habitability, The University of Texas, Austin Texas USA}

\author[0000-0002-9130-6747]{Eike W. Guenther}
\affiliation{Th\"uringer Landessternwarte Tautenburg, Sternwarte 5, 07778 Tautenburg, Germany}

\author[0000-0002-7031-7754]{Grzegorz Nowak}
\affiliation{Institute of Astronomy, Faculty of Physics, Astronomy and Informatics, Nicolaus Copernicus University, Grudzi\c{a}dzka 5, 87-100 Toru\'n, Poland\label{ia_ncu}}
\affiliation{Instituto de Astrofísica de Canarias (IAC), E-38205 La Laguna, Tenerife, Spain}
\affiliation{Departamento de Astrofísica, Universidad de La Laguna (ULL), E-38206 La Laguna, Tenerife, Spain}

\author[0000-0003-1257-5146]{Carina Persson}
\affiliation{Department of Space, Earth and Environment, Chalmers University of Technology, Onsala Space Observatory, SE-439 92 Onsala, Sweden}

\author{Phillip J. MacQueen}
\affiliation{McDonald Observatory, The University of Texas, Austin Texas USA}

\author[0000-0002-2086-3642]{Alexander J. Mustill}
\affiliation{Lund Observatory, Division of Astrophysics, Department of Physics, Lund University, Box 118, 22100 Lund, Sweden}

\author[0000-0001-6653-5487]{Juan Cabrera}
\affiliation{Institut f\"ur Planetenforschung, Deutsches Zentrum f\"ur Luft- und Raumfahrt, Rutherfordstr. 2, 12489 Berlin, Germany}

\author[0000-0001-9662-3496]{William D. Cochran}
\affiliation{McDonald Observatory, The University of Texas, Austin Texas USA}
\affiliation{Center for Planetary Systems Habitability, The University of Texas, Austin Texas USA}

\author[0000-0003-3742-1987]{Jorge~Lillo-Box} 
\affiliation{Centro de Astrobiologi\'ia (CAB), CSIC-INTA, Camino Bajo del Castillo s/n, ESAC campus, 28692, Villanueva de la Ca\~nada, Madrid, Spain}

\author[0000-0002-2696-1366]{David Hobbs}
\affiliation{Lund Observatory, Division of Astrophysics, Department of Physics, Lund University, Box 118, 22100 Lund, Sweden}

\author[0000-0001-9087-1245]{Felipe Murgas}
\affiliation{Instituto de Astrofísica de Canarias (IAC), E-38205 La Laguna, Tenerife, Spain}
\affiliation{Departamento de Astrofísica, Universidad de La Laguna (ULL), E-38206 La Laguna, Tenerife, Spain}

\author[0000-0002-0371-1647]{Michael Greklek-McKeon}
\affiliation{Division of Geological and Planetary Sciences, California Institute of Technology, Pasadena, CA, 91125, USA}

\author[0009-0006-3527-0424]{Hanna Kellermann}
\affiliation{University Observatory Munich, Faculty of Physics, Ludwig-Maximilians-Universit\"at München, Scheinerstr. 1, 81679 Munich, Germany}

\author{Guillaume H\'ebrard}
\affiliation{Institut d'astrophysique de Paris, UMR7095 CNRS, Universit\'e Pierre \& Marie Curie, 98bis boulevard Arago, 75014 Paris, France}
\affiliation{Observatoire de Haute-Provence, CNRS, Universit\'e d'Aix-Marseille, 04870 Saint-Michel-l'Observatoire, France}

\author[0000-0002-4909-5763]{Akihiko Fukui}
\affiliation{Komaba Institute for Science, The University of Tokyo, 3-8-1 Komaba, Meguro, Tokyo 153-8902, Japan}
\affiliation{Instituto de Astrofísica de Canarias (IAC), E-38205 La Laguna, Tenerife, Spain}

\author{Enric Pall\'e}
\affiliation{Instituto de Astrofísica de Canarias (IAC), E-38205 La Laguna, Tenerife, Spain}
\affiliation{Departamento de Astrofísica, Universidad de La Laguna (ULL), E-38206 La Laguna, Tenerife, Spain}

\author[0000-0002-4715-9460]{Jon M. Jenkins}
\affiliation{NASA Ames Research Center, Moffett Field, CA 94035, USA}

\author[0000-0002-6778-7552]{Joseph D. Twicken}
\affiliation{SETI Institute, Mountain View, CA 94043 USA}
\affiliation{NASA Ames Research Center, Moffett Field, CA 94035, USA}

\author[0000-0001-6588-9574]{Karen A. Collins}
\affiliation{Center for Astrophysics \textbar \ Harvard \& Smithsonian, 60 Garden Street, Cambridge, MA 02138, USA}

\author[0000-0002-8964-8377]{Samuel N. Quinn}
\affiliation{Center for Astrophysics \textbar \ Harvard \& Smithsonian, 60 Garden Street, Cambridge, MA 02138, USA}

\author{J\'an \v{S}ubjak}
\affiliation{Astronomical Institute, Czech Academy of Sciences, Fri{\v c}ova 298, 251 65, Ond\v{r}ejov, Czech Republic}
\affiliation{Center for Astrophysics \textbar \ Harvard \& Smithsonian, 60 Garden Street, Cambridge, MA 02138, USA}

\author[0000-0003-4745-2242]{Paul G. Beck}
\affiliation{Departamento de Astrofísica, Universidad de La Laguna (ULL), E-38206 La Laguna, Tenerife, Spain}
\affiliation{Instituto de Astrofísica de Canarias (IAC), E-38205 La Laguna, Tenerife, Spain}

\author[0000-0001-8627-9628]{Davide Gandolfi}
\affiliation{Dipartimento di Fisica, Universit\`a degli Studi di Torino, via Pietro Giuria 1, I-10125, Torino, Italy}

\author[0000-0002-0129-0316]{Savita Mathur}
\affiliation{Instituto de Astrofísica de Canarias (IAC), E-38205 La Laguna, Tenerife, Spain}
\affiliation{Departamento de Astrofísica, Universidad de La Laguna (ULL), E-38206 La Laguna, Tenerife, Spain}

\author[0000-0003-0047-4241]{Hans J. Deeg}
\affiliation{Instituto de Astrofísica de Canarias (IAC), E-38205 La Laguna, Tenerife, Spain}
\affiliation{Departamento de Astrofísica, Universidad de La Laguna (ULL), E-38206 La Laguna, Tenerife, Spain}

\author[0000-0001-9911-7388]{David~W.~Latham}
\affiliation{Center for Astrophysics \textbar \ Harvard \& Smithsonian, 60 Garden Street, Cambridge, MA 02138, USA}

\author[0000-0003-1762-8235]{Simon Albrecht}
\affiliation{Stellar Astrophysics Centre, Department of Physics and Astronomy, Aarhus University, Ny Munkegade 120, DK-8000 Aarhus C, Denmark}

\author[0000-0002-5971-9242]{David Barrado}
\affiliation{Centro de Astrobiologi\'ia (CAB), CSIC-INTA, Camino Bajo del Castillo s/n, ESAC campus, 28692, Villanueva de la Ca\~nada, Madrid, Spain}

\author{Isabelle Boisse}
\affiliation{Observatoire de Haute-Provence, CNRS, Universit\'e d'Aix-Marseille, 04870 Saint-Michel-l'Observatoire, France}
\affiliation{Aix Marseille Univ, CNRS, CNES, LAM, Marseille, France}

\author[0000-0002-7084-487X]{Herv\'e Bouy}
\affiliation{Laboratoire d'astrophysique de Bordeaux, Univ. Bordeaux, CNRS, B18N, all\'ee Geoffroy Saint-Hilaire, 33615 Pessac, France} 
\affiliation{Institut universitaire de France (IUF), 1 rue Descartes, 75231 Paris CEDEX 05}

\author{Xavier Delfosse}
\affiliation{Universit\'e Grenoble Alpes, CNRS, IPAG, 38000 Grenoble, France}

\author{Olivier Demangeon}
\affiliation{Instituto de Astrof{\'\i}sica e Ci\^encias do Espa\c{c}o, Universidade do Porto, CAUP, Rua das Estrelas, 4150-762 Porto, Portugal}

\author[0000-0002-8854-3776]{Rafael A. Garc\'{i}a}
\affil{Universit\'e Paris-Saclay, Universit\'e Paris Cit\'e, CEA, CNRS, AIM, 91191, Gif-sur-Yvette, France}

\author[0000-0002-3404-8358]{Artie P. Hatzes}
\affiliation{Th\"uringer Landessternwarte Tautenburg, Sternwarte 5, 07778 Tautenburg, Germany}

\author{Neda Heidari}
\affiliation{Institut d'astrophysique de Paris, UMR7095 CNRS, Universit\'e Pierre \& Marie Curie, 98bis boulevard Arago, 75014 Paris, France}

\author[0000-0002-5978-057X]{Kai Ikuta}
\affiliation{Department of Multi-Disciplinary Sciences, Graduate School of Arts and Sciences, The University of Tokyo, 3-8-1 Komaba, Meguro, Tokyo 153-8902, Japan}

\author{Petr Kab\'{a}th}
\affiliation{Astronomical Institute, Czech Academy of Sciences, Fri{\v c}ova 298, 251 65, Ond\v{r}ejov, Czech Republic}

\author[0000-0002-5375-4725]{Heather A. Knutson}
\affiliation{Division of Geological and Planetary Sciences, California Institute of Technology, Pasadena, CA, 91125, USA}

\author{John Livingston}
\affiliation{Astrobiology Center, 2-21-1 Osawa, Mitaka, Tokyo 181-8588, Japan}
\affiliation{National Astronomical Observatory of Japan, 2-21-1 Osawa, Mitaka, Tokyo 181-8588, Japan}
\affiliation{Astronomical Science Program, Graduate University for Advanced Studies, SOKENDAI, 2-21-1, Osawa, Mitaka, Tokyo, 181-8588, Japan}

\author[0000-0002-5084-168X]{Eder Martioli}
\affiliation{Laborat\'orio Nacional de Astrof\'{\i}sica, Rua Estados Unidos 154, 1006 37504-364, Itajubá - MG, Brazil}
\affiliation{Institut d'astrophysique de Paris, UMR7095 CNRS, Universit\'e Pierre \& Marie Curie, 98bis boulevard Arago, 75014 Paris, France}

\author[0000-0001-9526-9499]{Mar\'ia Morales-Calder\'on}
\affiliation{Centro de Astrobiologi\'ia (CAB), CSIC-INTA, Camino Bajo del Castillo s/n, ESAC campus, 28692, Villanueva de la Ca\~nada, Madrid, Spain}

\author{Giuseppe Morello}
\affiliation{Instituto de Astrofísica de Andalucía (IAA-CSIC), Gta. de la Astronomía s/n, 18008 Granada, Granada, Spain}
\affiliation{Instituto de Astrofísica de Canarias (IAC), E-38205 La Laguna, Tenerife, Spain}

\author[0000-0001-8511-2981]{Norio Narita}
\affiliation{Komaba Institute for Science, The University of Tokyo, 3-8-1 Komaba, Meguro, Tokyo 153-8902, Japan}
\affiliation{Astrobiology Center, 2-21-1 Osawa, Mitaka, Tokyo 181-8588, Japan}
\affiliation{Instituto de Astrofísica de Canarias (IAC), E-38205 La Laguna, Tenerife, Spain}

\author{Jaume Orell-Miquel}
\affiliation{Instituto de Astrofísica de Canarias (IAC), E-38205 La Laguna, Tenerife, Spain}
\affiliation{Departamento de Astrofísica, Universidad de La Laguna (ULL), E-38206 La Laguna, Tenerife, Spain}

\author{Hanna L. M. Osborne}
\affiliation{Mullard Space Science Laboratory, University College London, Holmbury St Mary, Dorking, Surrey RH5 6NT, UK}
\affiliation{European Southern Observatory, Karl-Schwarzschild-Straße 2, Garching bei München D-85748, Germany}

\author[0000-0002-6812-4443]{Dinil B. Palakkatharappil}
\affiliation{Universit\'e Paris-Saclay, Universit\'e Paris Cit\'e, CEA, CNRS, AIM, 91191, Gif-sur-Yvette, France}

\author{Viktoria Pinter}
\affiliation{Nordic Optical Telescope, Rambla José Ana Fernandez Pérez 7, 38711, Breña Baja, Spain} 
\affiliation{University of Craiova, Alexandru Ioan Cuza 13, 200585, Craiova, Romania}

\author[0000-0003-3786-3486]{Seth Redfield}
\affiliation{Astronomy Department and Van Vleck Observatory, Wesleyan University, Middletown, CT 06459, USA}

\author[0009-0009-5132-9520]{Howard M. Relles}
\affiliation{Center for Astrophysics \textbar \ Harvard \& Smithsonian, 60 Garden Street, Cambridge, MA 02138, USA}

\author[0000-0001-8227-1020]{Richard P. Schwarz}
\affiliation{Center for Astrophysics \textbar \ Harvard \& Smithsonian, 60 Garden Street, Cambridge, MA 02138, USA}

\author[0000-0002-6892-6948]{Sara Seager}
\affiliation{Department of Physics and Kavli Institute for Astrophysics and Space Research, Massachusetts Institute of Technology, Cambridge, MA 02139, USA}
\affiliation{Department of Earth, Atmospheric and Planetary Sciences, Massachusetts Institute of Technology, Cambridge, MA 02139, USA}
\affiliation{Department of Aeronautics and Astronautics, MIT, 77 Massachusetts Avenue, Cambridge, MA 02139, USA}

\author[0000-0002-1836-3120]{Avi Shporer}
\affiliation{Department of Physics and Kavli Institute for Astrophysics and Space Research, Massachusetts Institute of Technology, Cambridge, MA 02139, USA}

\author{Marek Skarka}
\affiliation{Astronomical Institute, Czech Academy of Sciences, Fri{\v c}ova 298, 251 65, Ond\v{r}ejov, Czech Republic}
\affiliation{Department of Theoretical Physics and Astrophysics, Faculty of Science, Masaryk University, Kotl\'a\v{r}sk\'a 267/2, 611~37 Brno, Czech Republic}

\author{Gregor Srdoc}
\affil{Kotizarovci Observatory, Sarsoni 90, 51216 Viskovo, Croatia}

\author[0000-0002-1812-8024]{Monika Stangret}
\affiliation{INAF – Osservatorio Astronomico di Padova, Vicolo dell'Osservatorio 5, 35122, Padova, Italy}

\author[0009-0006-1571-0306]{Luis Thomas}
\affiliation{University Observatory Munich, Faculty of Physics, Ludwig-Maximilians-Universit\"at München, Scheinerstr. 1, 81679 Munich, Germany}
\affiliation{Max-Planck Institute for Extraterrestrial Physics, Giessenbachstrasse 1, D-85748 Garching, Germany}

\author[0000-0001-5542-8870]{Vincent Van Eylen}
\affiliation{Mullard Space Science Laboratory, University College London, Holmbury St Mary, Dorking, Surrey RH5 6NT, UK}

\author[0000-0002-7522-8195]{Noriharu Watanabe}
\affiliation{Department of Multi-Disciplinary Sciences, Graduate School of Arts and Sciences, The University of Tokyo, 3-8-1 Komaba, Meguro, Tokyo 153-8902, Japan}

\author[0000-0002-4265-047X]{Joshua N. Winn}
\affiliation{Department of Astrophysical Sciences, Princeton University, Princeton, NJ 08544, USA}

\begin{abstract}
We report the discovery and characterization of a small planet, \planetc, on a 2.2-day orbit located interior to a previously known hot Jupiter, \planetb (P=\periodb, M=\massb, R=\radiusb) that exhibits grazing transits. The two planets are near 2:1 period commensurability, resulting in significant transit timing variations (TTVs) for both planets and transit duration variations (TDVs) for the inner planet. The TTV amplitude for \planetc is 15\% of the planet's orbital period, marking the largest TTV amplitude relative to the orbital period measured to date. Photodynamical modeling of ground-based radial velocity (RV) observations and transit light curves obtained with the Transiting Exoplanet Survey Satellite (\tess) and ground-based facilities leads to an inner planet radius of \radiusc and mass of \massc that locates the planet into the Sub-Neptune regime. The proximity to the 2:1 period commensurability leads to the libration of the resonant argument of the inner planet. The RV measurements support the existence of a third body with an orbital period of several thousand days. This discovery places the system among the rare systems featuring a hot Jupiter accompanied by an inner low-mass planet.
\end{abstract}


\section{Introduction} \label{sec:intro}

The discovery of the first exoplanet around a main sequence star, 51 Peg b, significantly advanced our understanding of planetary systems due to its dissimilarity (e.g. short orbital period) from any known planets in our Solar system \citep{1995Natur.378..355M}. Nearly 30 years later, the origins of such hot Jupiters (HJs)---gas giants with orbital periods less than 10 days and with planet masses higher than 0.1\mjup following the definition from \citet{2015ApJ...799..229W}---remain elusive. Theories suggest that HJs could form in situ \citep[e.g.,][]{2016ApJ...829..114B}, migrate inward from beyond the ice line through the interaction with the gas disk during formation \citep[e.g,][]{1996Natur.380..606L}, or undergo high-eccentricity migration \citep[HEM:][]{1996Sci...274..954R} at a later stage. For an overview, see \citet{2018ARA&A..56..175D} and references therein. 
 
The lack of detections of low-mass planets interior to HJs \citep{2012PNAS..109.7982S,2015ApJ...808...14M,2016ApJ...825...98H,2021AJ....162..263H} is a key argument supporting HEM as the dominant formation channel. However, exceptional systems where an inner low-mass planet accompanies an HJ have been detected, such as WASP-47 \citep{2012MNRAS.426..739H,2015ApJ...812L..18B,2023A&A...673A..42N}, WASP-84 \citep{2014MNRAS.445.1114A,2023MNRAS.525L..43M}, Kepler-730 \citep{2018RNAAS...2..160Z,2019ApJ...870L..17C}, TOI-2000 \citep{2023MNRAS.524.1113S}, WASP-132 \citep{2017MNRAS.465.3693H,2022AJ....164...13H,Grieves2024:2406.15986v1}, and TOI-1130 \citep{2020ApJ...892L...7H,2023A&A...675A.115K}. These systems cannot be explained by HEM and require dynamically quiet formation process, such as disk migration \citep{2005A&A...441..791F,2007A&A...472.1003F,2023AJ....165..171W,2023AJ....166..267W, 2024MNRAS.530.3934H}, or a less quiet process, such as in situ formation \citep{2021MNRAS.505.2500P}. 

In this paper, we present the discovery of another of these rare systems containing an HJ 
and a low-mass planet close to the 2:1 period commensurability, resulting in measurable TTVs, similar to TOI-1130. We report the discovery and characterization of a small planet, \planetc, interior to a known grazing hot Jupiter, \planetb, discovered using the Transiting Exoplanet Survey Satellite \citep[\tess;][]{2015JATIS...1a4003R} photometry and confirmed by \citet{2023MNRAS.526L.111G}. Our analysis also refines the orbital and geometric properties of \planetb.
 
\section{Observations} \label{sec:observations}

\subsection{Photometric Observations}
\label{subsec:photometry}

\subsubsection{TESS Photometry}
\label{subsubsec:tess}
TOI-1408 (TIC 364186197) was observed at 2-min cadence by \tess in Sectors~16, 17, 18, 19, 24, 25, 52, 57, 58, 59, 73, and 76 from 2019-09-12 to 2024-03-25, spanning 61 transits for \planetb and 114 transits for \planetc. We used the publicly available Presearch Data Conditioning (PDC) light curves \citep{2012PASP..124.1000S,2012PASP..124..985S,2014PASP..126..100S} produced by the Science Processing Operations Center \citep[SPOC:][]{2016SPIE.9913E..3EJ} at NASA Ames Research Center, downloaded from the Mikulski Archive for Space Telescopes.\!\footnote{\url{https://mast.stsci.edu}.}

\subsubsection{Ground-based Photometry}
\label{subsubsec:gb-phot}

\begin{table}[]
    \centering
  \caption{Ground-based photometry.}
\label{tab:photometry}
    \begin{tabular*}{\columnwidth}{@{\extracolsep{\fill}} rrrr}
    \toprule
    \toprule
time [BJD] & Flux & e\_Flux & Instrument \\
\midrule
2459501.794176 & 1.001 & 0.002 & M3\_g \\
2459501.794335 & 1.002 & 0.002 & M3\_g \\
2459501.794457 & 0.999 & 0.002 & M3\_g \\
2459501.794579 & 1.000 & 0.002 & M3\_g \\
2459501.794702 & 0.999 & 0.002 & M3\_g \\
2459501.794824 & 1.000 & 0.002 & M3\_g \\
\bottomrule
\end{tabular*}
\tablecomments{Table~\ref{tab:photometry} is published in its entirety in the machine-readable format. A portion is shown here for guidance regarding its form and content.}
\end{table}

\paragraph{NOT/ALFOSC}

We observed five \planetb transits in the $i$ band using the Alhambra Faint Object Spectrograph and Camera (ALFOSC) instrument installed at the 2.56-m Nordic Optical Telescope (NOT) at the Roque de los Muchachos Observatory on La Palma, Spain, between 2021-07-08 and 2022-08-28. The photometry was reduced with our pipeline following standard photometry practices \citep{Parviainen2019}. The reduced photometry from NOT and all the other instruments is available in Table~\ref{tab:photometry}.

\paragraph{LCOGT/Sinistro}

We observed six \planetb transits in the $i$ band using the Sinistro cameras installed at 1-m telescopes from the Las Cumbres Observatory Global Telescope \citep[LCOGT;][]{Brown:2013}  between 2021-07-21 and 2023-07-30. The photometry was reduced with the same pipeline as the NOT photometry. We also observed two transits of \planetb on 2021-05-21 and on 2022-09-11 in Pan-STARRS $z$-short band. The $z$-short band images were calibrated by the standard LCOGT {\tt BANZAI} pipeline \citep{McCully:2018} and differential photometric data were extracted using {\tt AstroImageJ} \citep{Collins:2017}.

\paragraph{LCOGT/MuSCAT3}

We observed two transit of \planetb using the multi-color MuSCAT3 instrument \citep{2020SPIE11447E..5KN} installed at the 2-m Faulkes Telescope North of LCOGT in Maui, Hawaii on 2021-10-14 and 2022-11-12. The transits were observed simultaneously in the $g$, $r$, $i$, and $z_s$ bands. The images were calibrated by the {\tt BANZAI} pipeline \citep{curtis_mccully_2018_1257560}, and aperture photometry was performed with a custom pipeline described in \citet{2011PASJ...63..287F}. We included only the first transit due to the low S/N of the second observation.  

\paragraph{Palomar/WIRC}

We observed one transit of \planetb on 2022-09-02 in the K-continuum band using the Wide-field Infrared Camera installed on the 5.1-m Hale Telescope at Palomar Observatory in California. 
We calibrated the images with the pipeline described in \cite{vissapragada2020}, and then performed aperture photometry and detrended the light curve with the procedure described in \cite{greklekmckeon2023}.

\subsection{Spectroscopic Observations}
\label{subsec:radial}

\begin{table}[]
    \centering
  \caption{RV measurements and their uncertainties.}
\label{tab:rvs}
    \begin{tabular*}{\columnwidth}{@{\extracolsep{\fill}} rrrr}
    \toprule
    \toprule
time [BJD] & RV [m/s] & e\_RV [m/s] & Instrument \\
\midrule
2459114.650183 & -33581.2 & 12.5 & McD HJST \\
2459115.814081 & -33383.9 & 19.8 & McD HJST \\
2459116.786824 & -33541.8 & 16.2 & McD HJST \\
2459133.725274 & -33413.8 & 19.9 & McD HJST \\
2459135.701365 & -33784.5 & 22.7 & McD HJST \\
2459143.690476 & -33627.5 & 8.7 & McD HJST \\
\bottomrule
\end{tabular*}
\tablecomments{Table~\ref{tab:rvs} is published in its entirety in the machine-readable format. A portion is shown here for guidance regarding its form and content.}
\end{table}

\paragraph{HARPS-N}

We obtained 60 high-resolution spectra using the High Accuracy Radial velocity Planet Searcher-North \citep[HARPS-N: $\lambda$\,$\in$\,(378--691)\,nm, $R\approx115\,000$;][]{2012SPIE.8446E..1VC} mounted at the 3.58-m Telescopio Nazionale Galileo (TNG) of Roque de los Muchachos Observatory in La Palma, Spain, under the observing program CAT21A\_119 between 2021-05-21 and 2024-06-19. The exposure time was set to 480--1800 seconds based on weather conditions and scheduling constraints, leading to a signal-to-noise ratio (SNR) per pixel of 36--112 at 550\,nm. The spectra were extracted using the off-line version of the HARPS-N DRS pipeline \citep{2014SPIE.9147E..8CC}, version  {\tt HARPN\_3.7}. Absolute radial velocities (RVs) were measured using an on-line version of the DRS, the YABI tool,\!\footnote{Available at \url{http://ia2-harps.oats.inaf.it:8000}.} by cross-correlating the extracted spectra with a G2 mask \citep{1996A&AS..119..373B}. 
The DRS RVs from HARPS-N and all other instruments can be found in Table~\ref{tab:rvs}.

\paragraph{McDonald}
We observed 89 high-precision spectra of \host using the McDonald Observatory 2.7-m Harlan J. Smith telescope (HJST) with its Tull Coud\'e spectrometer \citep[$\lambda$\,$\in$\,(340--1090)\,nm, $R\approx60\,000$;][]{1995PASP..107..251T}. We pass the starlight through an I$_2$ gas absorption cell in front of the spectrograph entrance slit in order to impose our high-precision RV metric on the spectrum. We used an exposure meter to terminate each observation when an SNR of about 25-40 per pixel was achieved.  All observations were reduced and 1-D spectra were extracted using standard IRAF routines. We computed the radial velocities given in Table~\ref{tab:rvs} using the AUSTRAL code \citep{2000A&A...362..585E}. 
\paragraph{TLS}
We obtained 232 spectra of \host using the Coud\'e-\'Echelle spectrograph of the 2-m-Alfred Jensch telescope of the Th\"uringer Landessternwarte (TLS) Tautenburg. The spectrograph's old camera (CCD3, $\lambda$\,$\in$\,(452--765)\,nm, $R=51\,000$) was updated to a new (Andor, $\lambda$\,$\in$\,(460--734)\,nm, $R=63\,000$) during the observing period. When modeling the RVs, we allowed for a velocity offset between the pre-upgrade and post-upgrade datasets because of the difference between the two cameras. We obtained 74 spectra with CCD3 and 158 spectra with the Andor-CCD with most exposure times of 1200~s. The observations were carried out using an iodine cell for wavelength calibrations. The RVs were obtained from the reduced spectra using {\tt viper} \citep{2021ascl.soft08006Z} and the co-added HARPS-N spectrum as a template.

\paragraph{CAFE}
We observed 20 spectra of \host with the Calar Alto Fiber-fed Echelle (CAFE) spectrograph \citep[$\lambda$\,$\in$\,(407--925)\,nm, $R=60\,000$;][]{2013A&A...552A..31A} mounted at the 2.2-m telescope of the Calar Alto observatory between 2021-11-10 and 2022-08-19 with a typical signal-to-noise ratio of 30. We used the observatory pipeline described in \cite{2020MNRAS.491.4496L} to perform the basic reduction and extraction of the spectra. This pipeline also determines the radial velocity by performing cross-correlation against a solar binary mask and correcting them from intra-night drifts using the Thorium-Argon frames obtained in between each science frame. Usually, several epochs were obtained per night with a mean individual radial velocity uncertainty of 17 m/s. The RVs were corrected for nightly zero points determined by observing the same sample of radial velocity standards every observing night.

\paragraph{SOPHIE}
We observed 15 high-precision spectra of \host with the SOPHIE \'echelle spectrograph ($\lambda$\,$\in$\,(387--694)\,nm, $R=75\,000$) between August 2020 and August 2023 installed at the 1.93-m telescope of Observatoire de Haute-Provence, France \citep{2008SPIE.7014E..0JP,2009A&A...505..853B}. For each spectrum, the exposure time was between 360~s and 1020~s, providing an SNR per pixel at 550~nm between 38 and 53 depending on the weather conditions. The radial velocities were extracted with the standard SOPHIE pipeline using cross-correlation as presented by \citet{2009A&A...505..853B} and refined by \citet{2024A&A...681A..55H}. The derived radial velocities have typical uncertainties of the order of $\pm7$~m/s.

\paragraph{MaHPS}
We observed 61 spectra between May and October 2022 using the 2.1 m Fraunhofer Telescope at the Wendelstein Observatory \rev{(WO)} in the German Alps. Our Manfred-Hirt-Planet Spectrograph (MaHPS) comprises the high-resolution spectrograph FOCES combined with the Menlo Systems Astrofrequency comb as a calibration light source ($\lambda$\,$\in$\,(380--880)\,nm, $R=60\,000$). For the data reduction from 2D to 1D, we used {\tt GAMSE}. For the comb calibration, creation of the b-spline optimized templates, and RV extraction via a fit, we used our pipeline {\tt MARMOT}. Descriptions of both can be found in \cite{HannaDoktorarbeit}. We provide two data sets one with Thar and one with comb calibration. 

\paragraph{TRES} We obtained 18 spectra with the Tillinghast Reflector Echelle Spectrograph \citep[TRES; $\lambda$\,$\in$\,(384--910)\,nm, $R=44\,000$;][]{tres} mounted on the 1.5-m Tillinghast Reflector at the Fred Lawrence Whipple Observatory on Mount Hopkins in southern Arizona between 2019-12-05 and 2024-06-17. The reduction and RV analysis followed the procedures described in \citet{2010ApJ...720.1118B} and \cite{2012ApJ...756L..33Q}. The main difference is that the template spectrum for the RV extraction was created by median-combining all of the out-of-transit observed spectra of \host (after shifting to align them).  Thus, these are relative velocities where the internal precision of each observation has been based on the scatter in the velocities between the individual echelle orders.  

\paragraph{OES}
We observed 14 spectra with the Ondrejov Echelle Spectrograph \citep[OES; $\lambda$\,$\in$\,(380--900)\,nm, $R=50\,000$;][]{2020PASP..132c5002K} installed on a 2-m Perek telescope in Ond\v{r}ejov, Czech
Republic. Exposure times varied from 2700~s to 3600~s depending on the weather conditions. The data are reduced via standard spectroscopic IRAF routines.
OES is used for the KESPRINT follow-up of TESS targets
\citep{2020AJ....159..151S,2022AJ....163..225T,2022MNRAS.513.5955K}.

\paragraph{HERMES}
We monitored the system for ten days in August 2020 with the High Efficiency and Resolution Mercator Echelle Spectrograph \citep[HERMES; $\lambda$\,$\in$\,(375--900)\,nm, $R=90\,000$;][]{2011A&A...526A..69R}, mounted on the 1.2~m Mercator telescope at the Spanish Observatorio del Roque de los Muchachos of the Instituto de Astrofísica de Canarias. We obtained these observations with simultaneous thorium-argon emission spectra, which allows a precision of 2 to 3 m/s \citep{2015A&A...573A.138B}.

\section{Analysis} \label{sec:analysis}

\subsection{Stellar characterization} \label{subsec:star}

We carried out a stellar analysis using the co-added HARPS-N spectra following an approach described in \citet{2017A&A...604A..16F} and \citet{2018A&A...618A..33P} using the empirical \href{https://github.com/samuelyeewl/specmatch-emp}{\tt {SpecMatch-Emp}} code \citep{2017ApJ...836...77Y} and the Spectroscopy Made Easy (\href{http://www.stsci.edu/~valenti/sme.html}{\tt{SME}}) analysis package  \citep{1996A&AS..118..595V, 2017A&A...597A..16P} version 5.22 to 
obtain the effective temperature, \teff, the logarithm of 
the surface gravity, \logg, \vsini, and the abundance of iron relative to hydrogen, \feh. The derived stellar parameters from {\tt{SME}} were used to model the stellar radius and mass with the spectral energy distribution Bayesian model averaging fitter 
\citep[\href{https://github.com/jvines/astroARIADNE}{\tt{ARIADNE};} ][]{2022MNRAS.513.2719V} using the 
{\tt {Phoenix~v2}} \citep{2013A&A...553A...6H}, {\tt {BtSettl}} \citep{2012RSPTA.370.2765A}, 
\citet{2003IAUS..210P.A20C}, and \citet{1979ApJS...40....1K} atmospheric model grids. The stellar rotation period, estimated from the spectroscopically derived \vsini and the stellar radius assuming that the star is seen equator-on, is \rot. This period is also identified from the \tess photometry with a Lomb-Scargle periodogram as well as by running the rotation pipeline \citep{2010A&A...511A..46M,2014A&A...568A..10G,2021ApJS..255...17S} on the Quick Look Pipeline \citep{2020RNAAS...4..206H} data, we derived a rotation period of $7.5\pm0.62$~d, indicating that the planetary orbital axis is indeed well-aligned with the spin of the star. This supports a more quiet formation scenario, such as disk migration. The results from the models are listed in Table~\ref{tab:stellar} alongside further stellar properties. 

\begin{table}
    \centering
    \caption{Stellar parameters of \host.}
    \begin{tabular*}{\columnwidth}{@{\extracolsep{\fill}} llr}
    \toprule
    \toprule
    Parameter & Value & Reference \\
    \midrule
    RA [J2000, epoch 2016] & 20:54:02.653 & 1 \\
    Dec [J2000, epoch 2016] & +72:34:50.34 & 1 \\
    parallax [mas] & $7.16\pm0.01$ & 1 \\
    \mua [\my] & $11.85\pm0.01$ & 1 \\
    \mud [\my] & $28.73\pm0.01$ & 1 \\
    RUWE & 0.88 & 1 \\
    Distance [pc] & $139.13\pm0.2$ & 1 \\
    V [mag] & $9.27\pm0.02$ & 2 \\
    J [mag] & $8.37\pm0.02$ & 3 \\
    Spectral type & F8\,V  & this work \\ 
    Age [Gyr] & $2.7\pm0.3$ & this work\\
    \teff [K] & $6117\pm31$ & this work\\
    \feh [dex] & $0.25\pm0.06$ & this work\\
    \logg [cm\,$\mathrm{s}^{-2}$] & $4.10\pm0.06$ & this work\\ 
    \vsini [\kms] & $9.8\pm0.7$ & this work \\ 
    \smass [\msun] & $1.31\pm0.01$& this work \\
    \sradius [\rsun] & $1.53\pm0.02$ & this work\\
    $\rho_\star$ [g\,$\mathrm{cm}^{-3}$] &  $0.51\pm0.02$ & this work\\
    \lum [$L_{\odot}$] & $2.96\pm0.10$ & this work \\  
    \bottomrule
    \bottomrule
    \end{tabular*}
    \tablerefs{1: Gaia DR3 \citep{2016A&A...595A...1G,2023A&A...674A...1G,2023A&A...674A..32B}, 2: Tycho-2 \citep{2000A&A...355L..27H}, 3: 2MASS \citep{2003yCat.2246....0C}.}
    \label{tab:stellar}
\end{table}

\subsection{Discovery of \planetc}
\label{subsec:discovery_of_planet_c}

A transit timing variation (TTV) search using  Python Tool for Transit Variation \citep[\pyttv;][]{2023A&A...675A.115K} revealed a clear TTV signal for \planetb, suggesting the presence of an additional body in the \host system. We used the Open Transit Search pipeline \citep[\href{https://github.com/hpparvi/opents}{\texttt{OpenTS;}}][]{Pope2016} to search for additional transit signals from the \tess Sectors 16, 17, 18, 19, 24, and 25. This analysis identified a transit-like signal with a period of 2.22~d, depth of 100~ppm, and transit duration of 0.62~h. This signal has not been reported as a TOI by the \tess mission. We also identified the transiting signal using the Détection Spécialisée de Transits code \citep[\texttt{DST;}][]{2012A&A...548A..44C}.

Following the detection of the tentative transit signal, we carried out a dynamical stability analysis using \rebound \citep{2012A&A...537A.128R}, which confirmed that the orbits are compatible with long-term stability and predicted large TTVs. Further analysis of individual \tess sectors supported the existence of large TTVs for \planetc. The transit durations and depths estimated from single sectors differed from those estimated by combining multiple sectors, leading to smearing of the transit shape. The orbital period was estimated to vary between 2.18 to 2.22~d from sector to sector. The individual transits have too low SNR to measure their transit durations, but we estimated the average transit durations combining two \tess sectors. Sectors 16 and 17 yield an average transit duration of 2.75~h, while Sectors 73 and 76 yield an average transit duration of 2.42~h. 

After our detection, the SPOC conducted an independent multi-sector transit search of the light curves up to Sector 59, using an adaptive, noise-compensating matched filter \citep{2002ApJ...575..493J,2010SPIE.7740E..0DJ,2020ksci.rept....9J}. The 4.42-day signal of \planetb\ was identified with a high significance, and after removing \planetb's transits, the next strongest signal ($8.16\sigma$) had an orbital period of $\sim2.22$ days. Despite exceeding the $7.1\sigma$ threshold, this detection was vetoed by the transit search pipeline's $\chi^2$ discriminator \citep{2013ApJS..206...25S}, likely due to smearing from large TTVs.

\subsection{Radial Velocities}
\label{subsec:Rvs}

We carried out a two-planet RV analysis using all the RV data described in Sect.~\ref{subsec:Rvs} with \pytransit's \texttt{RVLPF} RV modeling class \citep{Parviainen2015}. This code models the RV signals from two planets, incorporating an additional free RV offset and jitter term for each RV data set, a sinusoidal stellar rotational signal, and a quadratic trend in time to account for possible long-period companions. An analysis with wide normal priors on the planet b and c periods and transit centers identified with \texttt{OpenTS} leads to a posterior period estimate of $2.1613\pm0.0002$~d and an RV semi-amplitude of $5.0\pm1.0$~m/s that correspond to a minimum mass of $13.0\pm3.0$~\mearth. The RV information about \planetc comes from the HARPS-N data combining high precision with a long observing time span. For \planetb, we obtain a close-to-circular orbit with an RV semi-amplitude of $191.3\pm0.9$~m/s, translating to a minimum planet mass of $1.846\pm0.009$~\mjup.

The RV analysis also detected a clear non-linear trend showing evidence of an outer companion with an orbital period of thousands of days. Including the first two TRES points observed $\sim$200 days before the main observing campaign started gives a Keplerian orbit with a period of $\sim$2530~d with $K=195\pm4$\msunit, $M_\mathrm{p}=14.6\pm0.3$\mjup and an eccentricity of $0.35\pm0.02$. However, since the two points responsible for the solution are separated from the rest, we leave the characterization of the outer planet as future work requiring significantly longer RV follow-up.

Gaia DR3 astrometry indicates that the source is a primary star with 25 visibility periods and a ruwe of 0.88, indicating the solution is a good fit to linear space motion. At first sight, this would indicate it is a single star (Gaia astrometry could not detect planets with periods of a few days). However, the astrometry also shows an excess noise of 0.062 mas with a high significance of 4.7, which indicates there is a significant but small disturbance remaining. Gaia DR3 uses 34 months of data and may be sampling a substantial fraction of the suggested orbital period; however, very long orbital periods would require much more than 34 months of data to detect a companion.

\rev{We can estimate the astrometric signature of the long-period outer companion using the parameters given in this paper and assuming an almost circular orbit. The period can be used to estimate a semi-major axis for a roughly circular orbit, and the astrometric signature is obtained using 
\begin{equation}
\alpha = \left(\frac{M_p}{M_\star}\right)\left(\frac{a_p}{1~{\rm AU}}\right)\left(\frac{d}{1~{\rm pc}}\right)~\left[{\rm arsec}\right]\,.
\end{equation}
The suspected third planet gives an astrometric signature of 0.364 mas which should be easily detectable by Gaia, but with a period of almost 7 years compared to the Gaia mission duration of only $~$2.8 years, it could remain undetected until Gaia DR4 or DR5, especially if the planet is near apogee or if its period is underestimated.}

\subsection{Photodynamical analysis}
\label{subsec:photodyn}

\begin{figure*}[t!]
    \centering
    \includegraphics[width=\linewidth]{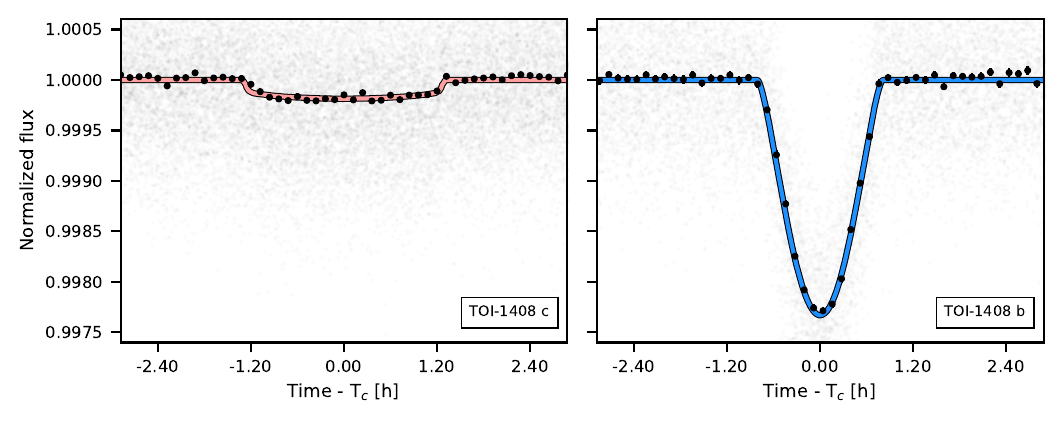}
    \includegraphics[width=\linewidth]{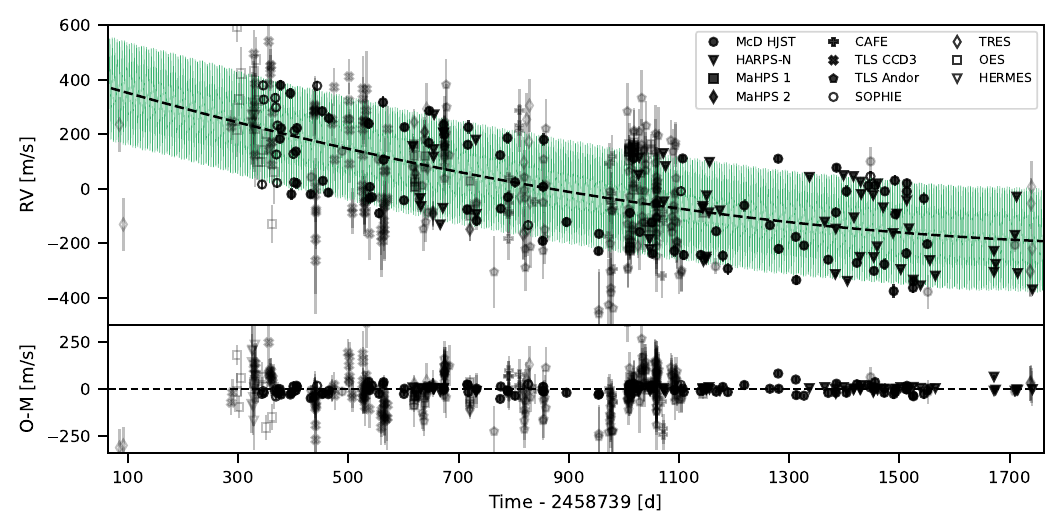}
    \caption{Photometry from \tess (top) and the RV measurements from different instruments (bottom) with the posterior models from the photodynamical analysis. The photometry is centered around the transit centers, with light gray points showing the raw photometry and black points with error bars show the photometry binned to 7~min. The RV figure shows the RV observations as black symbols with error bars, the photodynamical RV model as a green line, and the quadratic trend as a black dashed line. The gray symbols mark observations with uncertainties $>25$~m/s, and the black symbols observations with uncertainties $<25$~m/s. See Fig.~\ref{fig:gb_transits} in Appendix~\ref{sec:appendix.transits} for the ground-based photometry of \planetb and Fig.~\ref{fig:rv-mode-full} in Appendix~\ref{sec:appendix.rv_model} for a detailed illustration of the RV observations and the RV model.}
    \label{fig:data-and-models}
\end{figure*}

\begin{figure*}
    \centering
    \includegraphics[width=\linewidth]{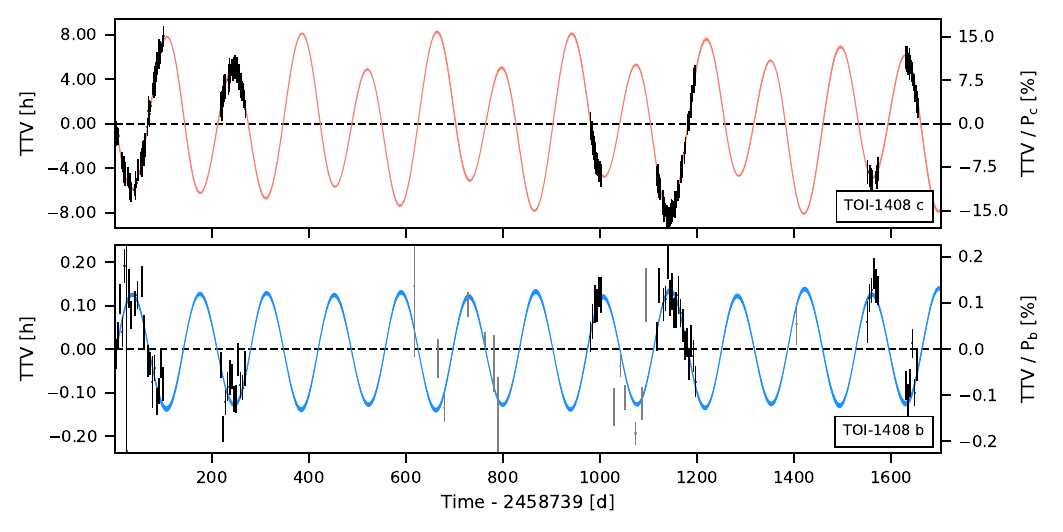}
    \caption{Posterior TTV model from the photodynamical modeling of \planetc (upper panel) and \planetb (lower panel). The colored lines span the 0.5 to 99.5 percentiles of the TTV model posterior distribution from the photodynamical analysis, but the uncertainties are mostly smaller than the line width.
    The dashed lines mark the subtracted mean orbital periods and 
    the TTVs with their individual uncertainties measured by fitting each transit center separately are shown as black (\tess) and gray (ground-based) points with error bars for comparison.}
    \label{fig:ttv-fit}
\end{figure*}

Since the planets strongly interact gravitationally, producing significant TTVs, we performed a photodynamical analysis by modeling the \tess photometry, the ground-based photometry, and the ground-based RVs from various facilities. This was done using \pyttv following the approach described in \citet{2023A&A...675A.115K}. We assumed a two-planet model, a sinusoidal RV signal to account for the stellar rotation based on the results from the stellar characterization in Sec.~\ref{subsec:star}, and a quadratic RV trend based on the results from the RV analysis in Sec.~\ref{subsec:Rvs}. The model is parametrized as described in \citet{2023A&A...675A.115K}, with the exception of the impact parameter, $b$. Instead, we used the grazing parameter, \gimp$=b/(1+k)$, where $k$ is the radius ratio, due to the grazing geometry of \planetb. The longitudes of the ascending notes, $\Omega$, are fixed to $\pi$ for both planets. The model is parameterized using the sampling parameters $\sqrt{e}\cos{\omega}$ and $\sqrt{e}\sin\omega$, but we also set a prior on the orbital eccentricities.  
The model parameters and their priors are listed in Table~\ref{tab:toi-1408_values}. 

The code models the photometry and RVs simultaneously using \rebound \citep{2012A&A...537A.128R, 2015MNRAS.446.1424R, 2020MNRAS.491.2885T} for dynamical integration, including General Relativity effects as implemented in \reboundx \citep{2020MNRAS.491.2885T}, and \pytransit \citep{Parviainen2015,2020MNRAS.499.1633P,2020MNRAS.499.3356P} for transit modeling. 
The analysis begins with a global optimization using the differential evolution method \citep{1997JGOpt..11..341S,Price2005}, followed by Markov Chain Monte Carlo (MCMC) sampling starting from the global optimization results with the \emcee sampler \citep{2013PASP..125..306F}.

We show the \tess photometry and RV measurements with their corresponding model in Fig.~\ref{fig:data-and-models}, the ground-based photometry in Fig.~\ref{fig:gb_transits} in Appendix~\ref{sec:appendix.transits}, the TTV model in Fig.~\ref{fig:ttv-fit}, and the full RV model in Fig.~\ref{fig:rv-mode-full} in Appendix \ref{sec:appendix.rv_model}. The posteriors for model parameters and the derived planetary parameters are listed in Table~\ref{tab:toi-1408_values}. The posterior mean ephemeris for \planetc is $T_{0,c}=2458739.845\pm0.002$~d and $P_c = 2.19592\pm0.00002$~d, while for \planetb, the ephemeris is $T_{0,b} = 2458740.8584\pm0.0002$~d and $P_{b} = 4.424703\pm0.000001$~d. \planetc exhibits a TTV period of 138~d with an amplitude of 8~h, and \planetb shows TTVs with an amplitude of 8~min. \planetc's TTV amplitude is 15\% of its orbital period, the largest TTV amplitude relative to a planet's orbital period known at the time of writing. The TDVs for \planetc calculated from the photodynamical model agree with the measured values. 

In a previous study by \citet{2023MNRAS.526L.111G}, \planetb was found to have a mass of $1.69\pm0.20$~\mjup and an eccentric orbit ($e=0.259\pm0.026$). While their mass value agrees within 1$\sigma$ with our (largely more precise) value of $1.86\pm0.02$~\mjup, we find the \planetb's orbit to be nearly circular. Additionally, our photodynamical modeling could constrain the \planetb's radius more reliably.  

\begin{table*}[!t]
\begin{threeparttable}
    \centering
    \small
    \caption{Photodynamical model parameters with their priors and posteriors, and the posteriors for derived planetary parameters. \NP{\mu, \sigma} stands for a normal prior with a mean $\mu$ and standard deviation $\sigma$, and \UP{a,b} stands for a uniform distribution from $a$ to $b$. The osculating orbital elements are valid for the reference time $\epoch=2458739.84$.}
    \label{tab:toi-1408_values}
    \begin{tabularx}{\textwidth}{@{\extracolsep{\fill}}lrrrr}
    \toprule
    \toprule
    \textit{Stellar parameter} & & Prior & Posterior & \\
    \midrule
    \sradius [\rsun] & & \NP{1.534,0.021} & $1.53\pm0.02$ & \\
    \smass [\msun] & & \NP{1.309,0.012} & $1.312\pm0.009$ & \\
    Limb darkening ${q}_1$ & & \UP{0,1} & $0.5\pm0.1$ & \\
    Limb darkening ${q}_2$ & & \UP{0,1} & $<0.85$\tablenotemark{a} & \\
    $\rho_\star$  [g\,$\mathrm{cm}^{-3}$] & & &  $0.52\pm0.02$ \\
    \\
    \textit{RV parameter} & & & & \\
     \midrule
    $\gamma$ [m\,$\mathrm{s}^{-1}$] & & $\mathcal{N}$\tablenotemark{b} & & \\
    $\dot\gamma$ [m\,$\mathrm{s}^{-1}$\,$\mathrm{days}^{-1}$] & & \NP{0,1} & $-0.344\pm0.002$ & \\
    $\ddot\gamma$ [m\,$\mathrm{s}^{-1}$\,$\mathrm{days}^{-2}$] & & \NP{0.000,0.001} & $0.000143\pm0.000004$ & \\
    \multicolumn{2}{l}{\hspace{-4pt}$\log_{10}$ RV-jitter [$\log_{10}$ m\,$\mathrm{s}^{-1}$]} & $\mathcal{N}$\tablenotemark{c} &  & \\
    $P_\mathrm{rot}$ [days] & & \NP{7.90,0.05} & $7.91\pm0.02$ & \\
    $A_\mathrm{rot}$ [m\,$\mathrm{s}^{-1}$] & & \UP{0,40} & $2.3\pm0.7$ & \\
    \\
     & \multicolumn{2}{c}{\planetc} &\multicolumn{2}{c}{\planetb}\\
    \textit{Fitted planet parameter} & Prior & Posterior & Prior & Posterior \\
    \midrule
    \p [days] & \NP{2.170,0.005} & $2.1664\pm0.0001$ & \NP{4.424,0.005} & $4.42587\pm0.00003$\\
    \epoch [BJD] & \NP{2458739.84, 0.01} & $2458739.847\pm0.004$ & \NP{2458740.85, 0.01} & $2458740.8581\pm0.0002$\\
    $\log_{10}$\,$M_\mathrm{p}$ [$\log_{10}\msun$] & \UP{-5.3, -3.9} & $-4.64\pm0.01$ & \NP{-2.7,0.03} & $-2.749\pm0.003$\\            
    \rpoverrstar & \NP{0.014,0.004} & $0.0134\pm0.0003$ & \UP{0.05,0.35} & $0.15\pm0.02$\\ 
    $\sqrt{e}\cos{\omega}$ & \UP{-0.5, 0.5} & $-0.353\pm0.0005$ & \UP{-0.5, 0.5} & $0.012\pm0.007$\\ 
    $\sqrt{e}\sin{\omega}$ & \UP{-0.5, 0.5} & $0.103\pm0.002$ & \UP{-0.5, 0.5} & $-0.046\pm0.006$\\      
    \gimp & \UP{0, 1} & $0.73\pm 0.02$ & \UP{0, 1} & $0.939\pm0.004$\\
    \\
    \textit{Derived planet parameter} & & & \\
    \midrule
    $M_\mathrm{p}$ [\mearth] &  & $7.6\pm0.2$ &  & $593\pm4$ \\  
    $R_\mathrm{p}$ [\rearth] &  & $2.22\pm0.06$ & & $25\pm4$\\   
    $\rho_\mathrm{p}$ [g\,$\mathrm{cm}^{-3}$] &  & $3.8\pm0.3$ & & $0.18^{+0.17}_{-0.08}$\\
    $\mathrm{T_{14}}$ [h] & & $2.67\pm0.06$ & & $1.65\pm0.03$\\
    $e$ & \NP{0,0.083} & $0.1353\pm0.0001$ & \NP{0,0.083} & $0.0023\pm0.0005$\\
    $\omega$ [$^\circ$] & & $286.3\pm0.3$ & & $170\pm10$\\  
    $i$ [$^\circ$] & & $82.6\pm0.3$ & & $82.4\pm0.2$\\        
    $a/R_\star$ & & $5.04\pm0.06$ & & $8.13\pm0.09$\\
    $a$ [AU] & & $0.03587\pm0.00008$ & & $0.05778\pm0.0001$\\
        \bottomrule
        \bottomrule
            \end{tabularx}
           \end{threeparttable}
        \tablenotetext{a}{The $q_2$ limb darkening coefficient is not well constrained so we give only its 99th percentile upper limit.}
        \tablenotetext{b}{The instrument-specific RV shifts, $\gamma$, have loosely informative priors based on the RV measurement medians and scatters.}
        \tablenotetext{c}{All RV jitter terms have the same prior \NP{-1,1}.}
\end{table*}

\section{Discussion} \label{sec:disucssion} 

\subsection{Dynamics}

Besides the large TTVs, \planetc also exhibits transit duration variations (TDVs). Forward modeling of the posterior solution from Table~\ref{tab:toi-1408_values} allowed us to identify three main contributions to the TDVs, similar to those observed in KOI-142 \citep{2013ApJ...777....3N}. The largest contribution arises from the variability in $e\cos{\omega}$ with a period of $\sim 1$~y and a peak-to-peak (ptp) amplitude of 0.4~h. The second largest contribution is due to variations in the orbital inclination, with a period of $\sim 15$~y and a ptp amplitude of 0.2~h. The third contribution comes from the variation in tangential velocity at mid-transit, which has a period similar to the TTV period and a ptp amplitude of 0.08~h. 

\begin{figure*}
    \includegraphics[width=0.505\textwidth]{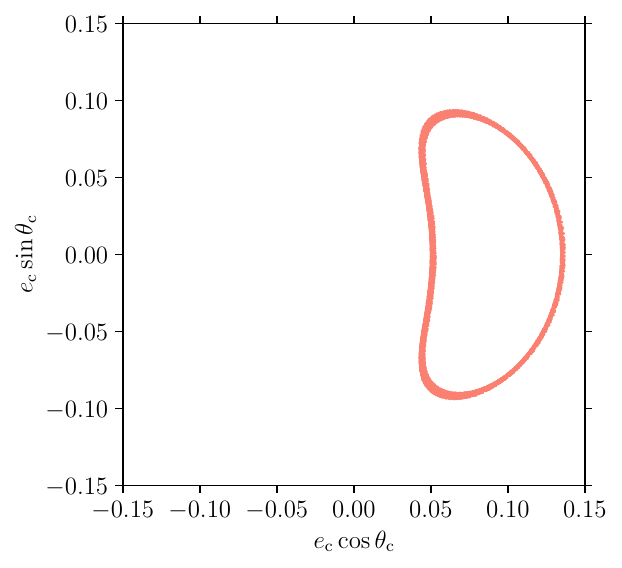}
    \includegraphics[width=0.495\textwidth]{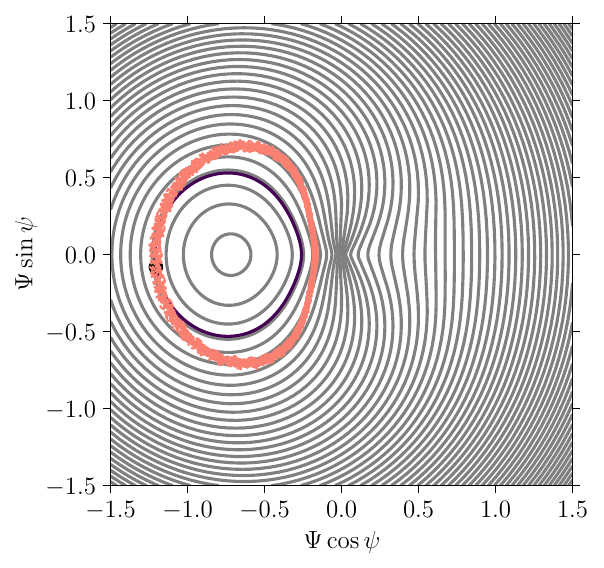}
    \caption{Resonant behavior of the \host b--c pair. Left: behavior of the single-planet resonant argument, $\theta_\mathrm{c}$, from the photodynamical simulations, showing libration about zero. Right: the same solution, transformed to the canonical resonance model of \cite{2016ApJ...823...72N}, where $\Psi$ is a function of both planets' eccentricities, and $\psi$ of both planets' resonant arguments, although both are dominated by the inner planet c. This numerical solution is marked in salmon, and $\psi$ librates about $\pi$. However, a comparison to level curves of the Hamiltonian (grey, with black marking the analytical curve corresponding to the numerical solution) shows that no resonant separatrix exists for these system parameters.}
    \label{fig:resonance}
\end{figure*}

The two planets of \host lie close to the 2:1 period commensurability ($P_\mathrm{b}/P_\mathrm{c}\approx2.04$), raising the question of whether the system is dynamically in resonance. We first checked for libration of the resonant argument of the lower-mass interior planet $\theta_\mathrm{c}=2\lambda_\mathrm{b}-\lambda_\mathrm{c}-\varpi_\mathrm{c}$. This resonant argument shows libration about $0^\circ$, suggesting a resonant state (Fig.~\ref{fig:resonance}, left). To understand the resonant dynamics more deeply, we then compared the system's behavior to the Hamiltonian resonance model of \cite{2016ApJ...823...72N}. The resonant argument in this model still librates (about $180^\circ$), but when we examine the full phase portrait, the resonant separatrix does not exist (Fig.~\ref{fig:resonance}, right). The separation to resonance, quantified by the parameter $\delta$ of \cite{2016ApJ...823...72N}, is around $\delta=0.3$ for all posterior draws (mean $\delta = 0.297$ with a standard deviation of $0.002$). The resonant separatrix exists only for $\delta\ge1$, and hence this system is by this definition just wide of resonance, similar to KOI-142 and TOI-2202 \citep{2022ApJ...925...38N}. This could imply that convergent migration in the protoplanetary disc did not proceed far enough to drive the system deeply into resonance, or alternatively, that tidal forces or other effects caused the orbits to diverge and escape from the resonant state \citep{2012A&A...546A..71D}.

\subsection{Stability}
\label{sec:stab}

\begin{figure*}
    \centering    
    \includegraphics[width=\linewidth]{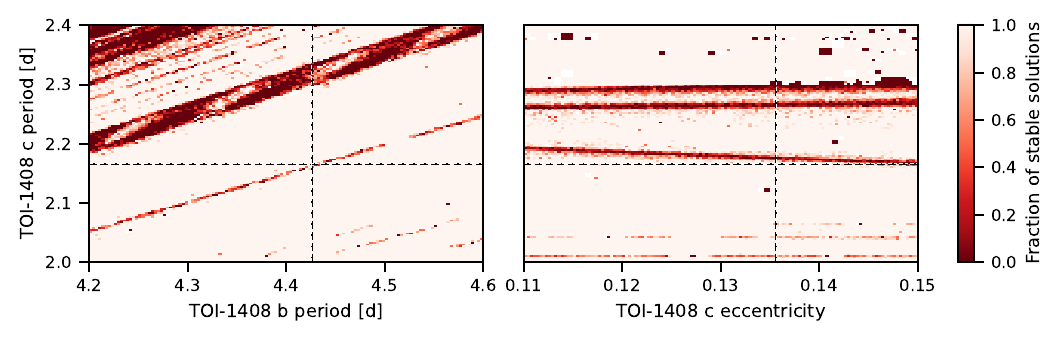}
    \caption{Two-dimensional MEGNO maps probing the stability near the posterior solution from the photodynamical analysis. The color indicates the orbital stability, and the dashed lines show the parameter posterior median values listed in Table~\ref{tab:toi-1408_values}.}
    \label{fig:megno}
\end{figure*}

We performed a numerical stability analysis using \rebound and its implementation of the Mean Exponential Growth factor of Nearby Orbits \citep[MEGNO;][]{2000A&AS..147..205C} indicator to determine if the planetary system lies in a stable configuration, which we define as $|\text{MEGNO} - 2| < 0.4$. We mapped the system's stability in the $\p_\mathrm{b}-\p_\mathrm{c}$, and $e_\mathrm{c}-\p_\mathrm{c}$ parameter spaces by drawing samples from the photodynamical model posterior and replacing the mapped parameters with samples created using a quad-tree based importance sampler. We simulated the system for 1.6 million \planetc orbits for each sample using the WHFast integrator \citep{2015MNRAS.452..376R} and saved the MEGNO indicator. We visualize the results as a stability map shown in Fig.~\ref{fig:megno}, where the value of a cell represents the fraction of stable orbits within that cell. It is worth noting that the system is close to a narrow zone of instability.

\subsection{Implications from Occurrence rates}

The results highlight the unique position of \host as a stable island in a chaotic environment, particularly given the low occurrence rate of inner companions to HJs, estimated at $1.4\%\pm1.0\%$ by \citet{2024MNRAS.530.3934H}. Their findings, based on N-body simulations spanning a period of $10^{8}$~years, agree with the lower limit from a previous study by \citet{2023AJ....165..171W}, which reported a rate of non-aligned nearby planetary companion to hot Jupiters at $12\%\pm6\%$ based on a search for TTVs in the Kepler sample. \citet{2024MNRAS.530.3934H} also found that the occurrence rate of inner companions to HJ significantly decreases as the HJ's orbital period shortens, resulting in only a few stable systems with HJ orbital periods less than 6~d. 

\section{Conclusions}
In this study, we report the discovery and photodynamical characterization of \planetc, a transiting sub-Neptune on a 2.2-day orbit located interior to a transiting hot Jupiter, \planetb (P=\periodb, M=\massb, R=\radiusb). The planets are near a 2:1 period commensurability with librating resonant arguments, yet both remain outside the resonant configuration. This configuration leads to remarkable TTVs and TDVs, with \planetc exhibiting the largest TTV amplitude relative to its orbital period recorded thus far.  

The existence of a small inner planet in such a tight orbit around a hot Jupiter adds a valuable data point against the backdrop of current planet formation theories, challenging the typical scenarios suggested for close-in giant planets. The planets in the \host system are both transiting and exhibit measurable TTVs, similar to TOI-1130, but in a much tighter orbit configuration. 

Additionally, RV measurements suggest a third, long-period outer body in the system, indicating a complex dynamical environment, such as WASP-47 and WASP-132. The system's dynamic nature and the likely presence of a third body invite further observational campaigns to refine the orbital parameters and investigate long-term stability.

This research not only deepens our understanding of multi-planet systems involving hot Jupiters with inner low-mass companions but also underscores the need for continued exploration to uncover the diverse architectures of exoplanetary systems.

\section{Acknowledgments}
\rev{We thank the anonymous referee for their timely and positive feedback.} This work was supported by the KESPRINT collaboration, an international consortium devoted to the characterization and research of exoplanets discovered with space-based missions (\url{https://kesprint.science/}). This paper includes data collected with the TESS mission, obtained from the MAST data archive at the Space Telescope Science Institute (STScI). Funding for the TESS mission is provided by the NASA Explorer Program. STScI is operated by the Association of Universities for Research in Astronomy, Inc., under NASA contract NAS 5–26555. We acknowledge the use of public TESS Alert data from pipelines at the TESS Science Office and at the TESS Science Processing Operations Center. Resources supporting this work were provided by the NASA High-End Computing (HEC) Program through the NASA Advanced Supercomputing (NAS) Division at Ames Research Center for the production of the SPOC data products. Based on observations made with the Nordic Optical Telescope, owned in collaboration by the University of Turku and Aarhus University, and operated jointly by Aarhus University, the University of Turku and the University of Oslo, representing Denmark, Finland and Norway, the University of Iceland and Stockholm University at the Observatorio del Roque de los Muchachos, La Palma, Spain, of the Instituto de Astrofisica de Canarias. The data presented here were obtained [in part] with ALFOSC, which is provided by the Instituto de Astrofisica de Andalucia (IAA) under a joint agreement with the University of Copenhagen and NOT. This article is based on observations made with the LCOGT telescopes, one of whose nodes is located at the Observatorios de Canarias del IAC on the island of Tenerife in the Observatorio del Teide. This paper is based on observations made with the MuSCAT3 instrument, developed by the Astrobiology Center and under financial supports by JSPS KAKENHI (JP18H05439) and JST PRESTO (JPMJPR1775), at Faulkes Telescope North on Maui, HI, operated by the Las Cumbres Observatory. \rev{This research has made use of the Exoplanet Follow-up Observation Program (ExoFOP; DOI: 10.26134/ExoFOP5) website, which is operated by the California Institute of Technology, under contract with the National Aeronautics and Space Administration under the Exoplanet Exploration Program}. This paper includes data taken at The McDonald Observatory of The University of Texas at Austin. Based on observations made with the Italian Telescopio Nazionale Galileo (TNG) operated on the island of La Palma by the Fundaci\'on Galileo Galilei of the INAF (Istituto Nazionale di Astrofisica) at the Spanish Observatorio del Roque de los Muchachos of the Instituto de Astrofisica de Canarias under programmes CAT19A\_162, ITP19\_1 and A41TAC\_49. Based partly on observations collected at the Centro Astronómico Hispano en Andalucía (CAHA) at Calar Alto, operated jointly by Junta de Andalucía and Consejo Superior de Investigaciones Científicas (IAA-CSIC). This work has made use of data from the European Space Agency (ESA) mission
{\it Gaia} (\url{https://www.cosmos.esa.int/gaia}), processed by the {\it Gaia}
Data Processing and Analysis Consortium (DPAC,
\url{https://www.cosmos.esa.int/web/gaia/dpac/consortium}). Funding for the DPAC
has been provided by national institutions, in particular the institutions
participating in the {\it Gaia} Multilateral Agreement
J.~K. gratefully acknowledges the support of the Swedish National Space Agency (SNSA; DNR 2020-00104) and of the Swedish Research Council  (VR: Etableringsbidrag 2017-04945)
H.P. acknowledges support from the Spanish Ministry of Science and Innovation with the Ramon y Cajal fellowship number RYC2021-031798-I, and funding from the University of La Laguna and the Spanish Ministry of Universities.
AJM acknowledges financial support from the Swedish National Space Agency (Career Grant 2023-00146). We acknowledge financial support from the Agencia Estatal de Investigaci\'on of the Ministerio de Ciencia e Innovaci\'on (MCIN/AEI, DOI: 10.13039/501100011033) and the ERDF “A way of making Europe” through project PID2021-125627OB-C32, and from the Centre of Excellence “Severo Ochoa” award to the Instituto de Astrofisica de Canarias. Also from MCIN/AEI, HJD acknowledges funding under grant PID2019-107061GB-C66 and G. M. under the Severo Ochoa grant CEX2021-001131-S and the Ram\'on y Cajal grant RYC2022-037854-I.
J.L-B. acknowledges financial support received from "la Caixa" Foundation (ID 100010434) and from the European Unions Horizon 2020 research and innovation programme under the Marie Slodowska-Curie grant agreement No 847648, with fellowship code LCF/BQ/PI20/11760023. This research has also been partly funded by the Spanish State Research Agency (AEI) Projects No. PID2019-107061GB-C61 and MDM-2017-0737 Unidad de Excelencia “María de Maeztu”- Centro de Astrobiología, CSIC/INTA. G.N. thanks for the research funding from the Ministry of Education and Science programme the "Excellence Initiative - Research University" conducted at the Centre of Excellence in Astrophysics and Astrochemistry of the Nicolaus Copernicus University in Toru\'n, Poland.
This work is partly supported by JSPS KAKENHI Grant Numbers JP24H00017, JP24H00248, JP24K00689, JP24K17082 and JSPS Bilateral Program Number JPJSBP120249910.
P.G.B. acknowledges support by the Spanish Ministry of Science and Innovation with the \textit{Ram{\'o}n\,y\,Cajal} fellowship number RYC-2021-033137-I and the number MRR4032204.
E.M. acknowledges funding from FAPEMIG under project number APQ-02493-22 and a research productivity grant number 309829/2022-4 awarded by CNPq, Brazil.
D.G. gratefully acknowledges the financial support from the grant for internationalization (GAND\_GFI\_23\_01) provided by the University of Turin.
M.S. acknowledges the support of the Italian National Institute of Astrophysics (INAF) through the project 'The HOT-ATMOS Project: characterizing the atmospheres of hot giant planets as a key to understand the exoplanet diversity' (1.05.01.85.04).
J.\v{S}. acknowledges the support from GACR grant 23-06384O. PK and MS are thankful for the support from grant LTT-20015.
P.C. acknowledges the DFG through grant CH 2636/1-1 and the support of the Department of Atomic Energy, Government
of India.
E.W.G acknowledges support from the Th\"uringer Ministerium f\"ur Wirtschaft, Wissenschaft und Digitale Gesellschaft.
R.A.G. and D.B.P. acknowledge the support from PLATO CNES grant. S.M.\ acknowledges support by the Spanish Ministry of Science and Innovation with the grant no. PID2019-107061GB-C66 and through AEI under the Severo Ochoa Centres of Excellence Programme 2020--2023 (CEX2019-000920-S). 


%

\vspace{5mm}
\facilities{NOT, CTIO:1.3m,
CTIO:1.5m, CXO, LCOGT, Hale, TNG, Smith, TLS, CA0:2.2m, OHP:1.93m, WO:2m, FLWO:1.5m, OO:2, Mercator1.2m.}


\software{PyTTV \citep{2023A&A...675A.115K},
          PyTransit \citep{Parviainen2015}, Rebound \citep{2012A&A...537A.128R},
          LDTk \citep{Parviainen2015b}, emcee \citep{2013PASP..125..306F},
          AstroImageJ \citep{Collins:2017}, 
          TAPIR \citep{Jensen:2013},
          Astropy \citep{2013A&A...558A..33A,2018AJ....156..123A,2022ApJ...935..167A},
          NumPy \citep{VanderWalt2011}, SciPy \citep{2020SciPy-NMeth}, matplotlib \citep{Hunter2007}
          }

\appendix

\section{Ground-based transits for \planetb}
\label{sec:appendix.transits}
\begin{figure*}
    \centering
    \includegraphics{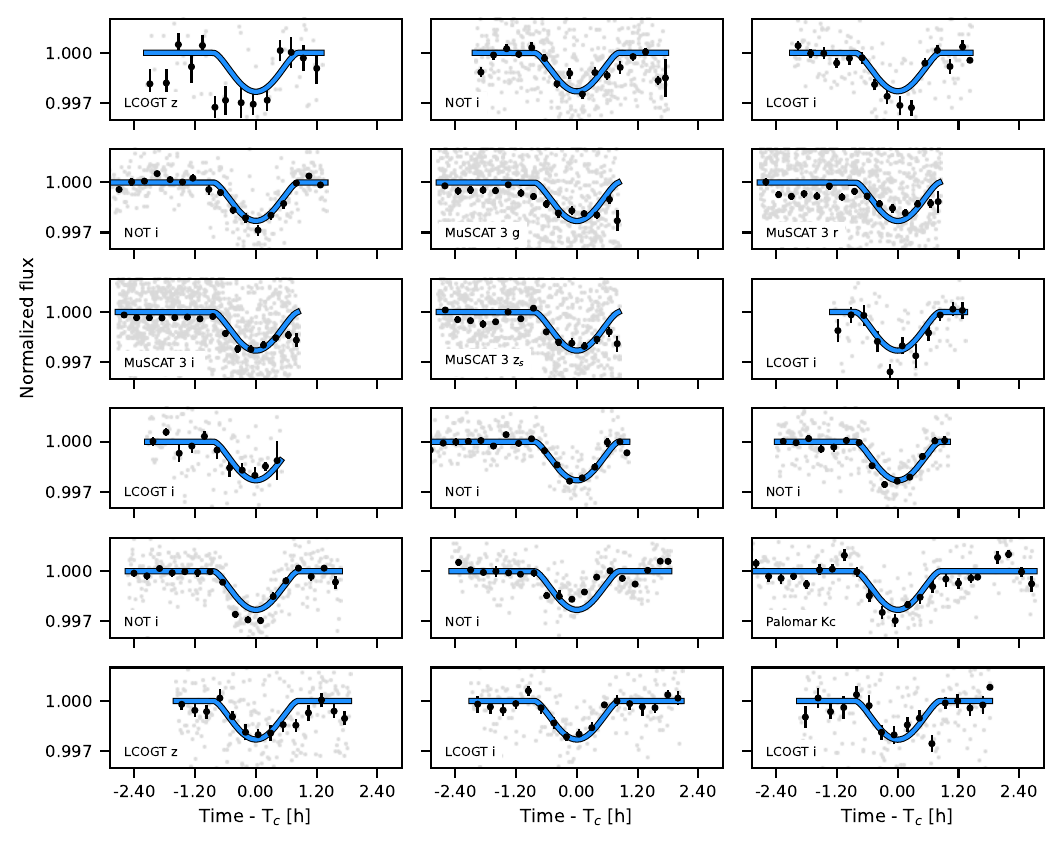}
    \caption{Ground-based transit measurements observed with the different facilities shown as gray points, the same points binned to 15-min as black points with error bars, and the photodynamical model in blue.}
    \label{fig:gb_transits}
\end{figure*}

\section{Radial velocity model from the photodynamical analysis}
\label{sec:appendix.rv_model}
\begin{figure*}
    \centering
    \includegraphics[width=\linewidth]{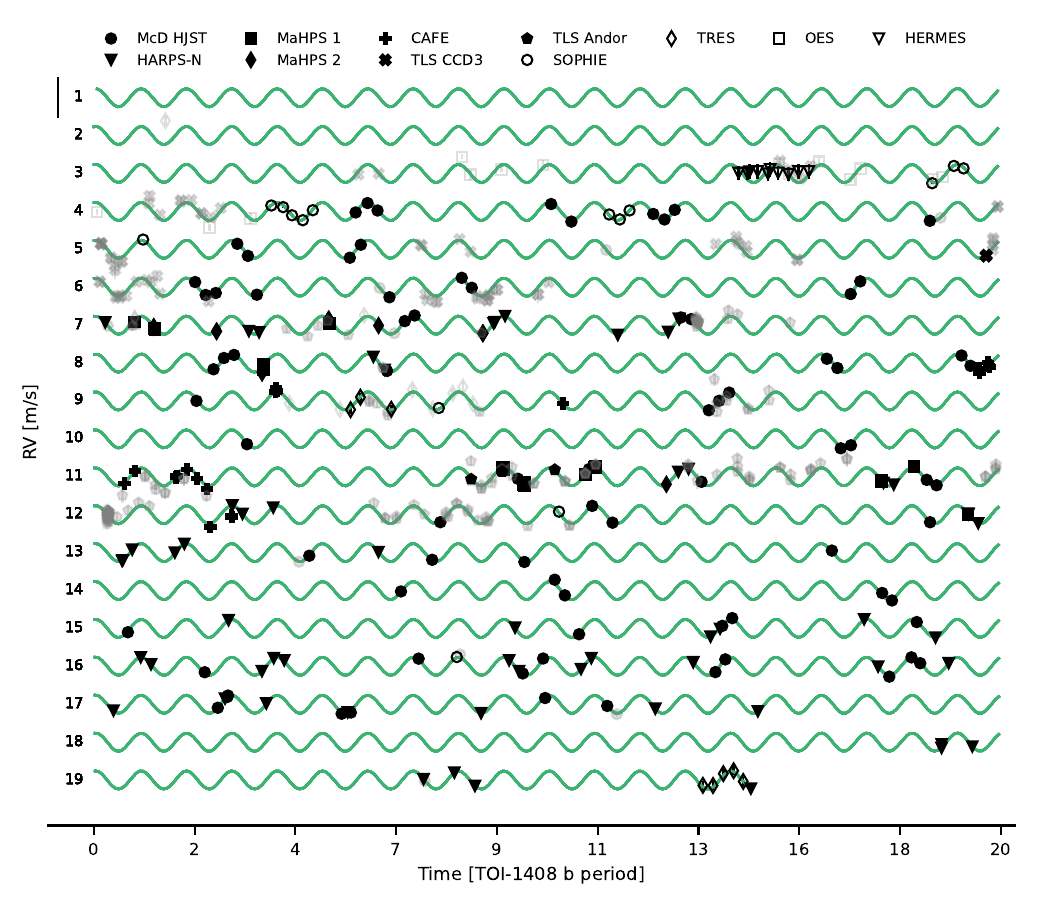}
    \caption{Radial velocity observations together with the model from the photodynamical analysis. One line covers 20 \planetb periods, after which the data and the model are offset down by 800~m/s. The vertical bar at line 1 shows the scale, spanning the RV range from -400 to 400 m/s. The light gray symbols mark RV observations with uncertainties larger than 25~m/s, and the black symbols observations with uncertainties smaller than 25~m/s.}
    \label{fig:rv-mode-full}
\end{figure*}

\bibliography{toi-1408}{}
\bibliographystyle{aasjournal}

\end{document}